\documentclass[%
 reprint, 
showpacs,
 amsmath,amssymb,
 aps,
prb,
floatfix,
 lengthcheck,
]{revtex4-1}

\usepackage{amssymb,amsmath}

\usepackage[caption=false]{subfig} 
\usepackage{graphicx}

\newcommand{\myL}[1]{{\cal L}^{(#1)}}

\DeclareMathAlphabet{\gcal}{OMS}{cmsy}{m}{n}
\newcommand{\myT}{{\gcal T}}

\newlength{\mycolumnwidth} 
\begin{document}
\author{J.~P.~Bergfield}
\affiliation{College of Optical Sciences, University of Arizona, 1630 East University Boulevard, AZ 85721}
\email{justinb@email.arizona.edu}
\author{M.~Solis}
\author{C.~A.~Stafford}
\affiliation{Department of Physics, University of Arizona, 1118 East Fourth Street, Tucson, AZ 85721}

\title{Giant Thermoelectric Effect from Transmission Supernodes}


\date{\today}

\begin{abstract}
We predict an enormous order-dependent quantum enhancement of thermoelectric effects in the vicinity of a higher-order `supernode' in the transmission spectrum of a nanoscale junction.  Single-molecule junctions based on 3,3'-biphenyl and polyphenyl ether (PPE) are investigated in detail.  The nonequilibrium thermodynamic efficiency and power output of a thermoelectric heat engine based on a 1,3-benzene junction are calculated using many-body theory, and compared to the predictions of the figure-of-merit $ZT$.  
\end{abstract}

\maketitle


Thermoelectric (TE) devices are highly desirable since they can directly convert between thermal and electrical energy.  Electrical power can be supplied to such a device to either heat or cool adjoining reservoirs (Peltier effect) or alternatively, the flow of heat (e.g. from a factory or car exhaust) can be converted into usable electrical power (Seebeck effect).  Often, the efficiency of a TE device is characterized by the dimensionless figure-of-merit $ZT$=$S^2GT/\kappa$, constructed with the rationale that an efficient TE device should simultaneously:  maximize the electrical conductance $G$ so that current can flow without much Joule heating,  minimize the thermal conductance $\kappa$ in order to maintain a temperature gradient across the device, and maximize the Seebeck coefficient $S$ to ensure that the coupling between the electronic and thermal currents is as large as possible.\cite{Bell08,DiSalvo99}  Generally, however, $ZT$ is difficult to maximize because these properties are {\em highly correlated} with one another,\cite{Hochbaum08, Majumbdar04, Snyder08} a fact that becomes more pronounced at the nanoscale where the number of degrees of freedom available is small. 

If a TE material were found exhibiting $ZT$$\geq$4 it would constitute a commercially viable solution for many heating and cooling problems at both the macro- and nano-scales, with no operational carbon footprint.\cite{DiSalvo99}  Currently, the best TE materials available in the laboratory exhibit $ZT$$\approx$3, whereas for commercially available TE {\em devices} $ZT$$\approx$1, owing to various packaging and fabrication challenges.\cite{Bell08,Harman02}  

In a previous article, enhanced thermoelectric effects were found in the vicinity of a transmission node of a quantum tunneling device.  Generically, the transmission probability vanishes quadratically as a function of energy at such a transmission node.\cite{Bergfield09b}  Here we present results for a class of two-terminal single-molecule junctions (SMJ) with higher-order `supernodes' in their transmission spectra.  In the vicinity of a 2$n$$^{\rm th}$ order supernode:
\begin{equation}
\myT(E) \propto (E-\mu_{\rm node})^{2n},
	\label{eq:T_supernode}
\end{equation}
where $\mu_{\rm node}$ is the energy of the node.  We find that junctions possessing such supernodes exhibit a scalable order-dependent quantum-enhanced thermoelectric response.

\begin{figure}[b]
	\centering
		\includegraphics[width=\mycolumnwidth]{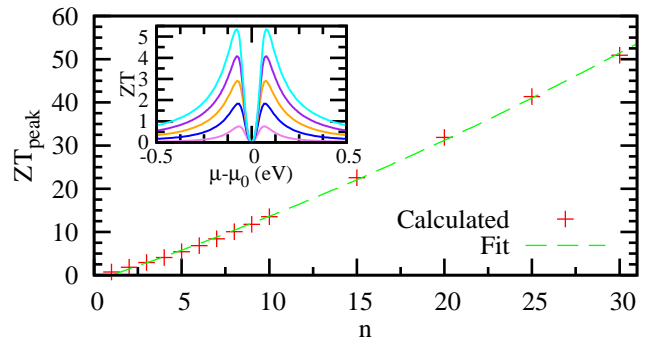}
		\caption{Near a 2$n$$^{\rm th}$ order {\em supernode} in a device's transmission spectrum, we find an order-dependent enhancement of the thermoelectric response which is limited only by the electronic coherence length.  Calculations were performed for a polyphenyl ether (PPE) SMJ with $n$ repeated phenyl groups at room temperature ($T$=300K) with $\Gamma$=0.5eV. Notice that the enhancement is super-linear in $n$.  Inset: $ZT$ as a function $\mu$ for n=$1\ldots5$.}
\label{fig:ZT_vs_n}
\end{figure}

As an example, $ZT$ of a supernode-possessing polyphenyl ether (PPE)-based SMJ is shown as a function of repeated phenyl unit number $n$ in Fig.~(\ref{fig:ZT_vs_n}).  As illustrated in the figure, $ZT_{\rm peak}$ scales super-linearly in $n$ whereby $ZT_{\rm peak}$=4.1 in a junction composed of just four phenyl groups ($n$=4).  Although we focus on molecular junctions in this article, it should be stressed that our results are applicable to any device with transmission nodes arising from coherent electronic transport.

As an engineering rule-of-thumb, $ZT$ has been widely used to characterize the bulk thermoelectric response of materials.\cite{Bell08,DiSalvo99,Snyder08}  At the nanoscale, however, it is unclear the extent to which $ZT$ is applicable, since bulk scaling relations for transport may break down due to quantum effects.\cite{Datta95}  Moreover, $ZT$ is a linear response metric, and cannot {\em a priori} predict nonequilibrium thermoelectric response.  

We investigate the efficacy of $ZT$ as a predictor of nonequilibrium device {\em performance} at the nanoscale by calculating the thermodynamic efficiency and power of an interacting quantum system using both nonequilibrium many-body\cite{Bergfield09} and H\"uckel theories.  We discover that in both theories, variations of $ZT$ and thermodynamic efficiency are in good qualitative agreement. However, large discrepancies between thermoelectric effects calculated within many-body and H\"uckel theory are found in the resonant tunneling regime, indicating the essential role of electron-electron interactions in nanoscale thermoelectricity.  For a thermoelectric quantum tunneling device, we find that the power output can be changed significantly by varying an external parameter, such
as a gate voltage, and that this variation is {\it not correlated} with the variation of $ZT$.

%

Neglecting inelastic processes, which are strongly suppressed at room temperature in SMJs, the current flowing into lead $1$ of a two-terminal junction may be written as follows:\cite{Bergfield09b}
\begin{equation}
\label{eq:Iq_ButtikerForm}
I^{(\nu)}_1=\frac{1}{h} \int_{-\infty}^\infty dE\; (E-\mu_1)^\nu \,\myT(E)\left[f_2(E)-f_1(E)\right],
\end{equation}
where $\nu=0$ ($\nu$=1) for the number (heat) current, $f_\alpha(E)$ is the Fermi function for lead $\alpha$ with chemical potential $\mu_\alpha$ and inverse temperature $\beta_\alpha$, and $\myT(E)$ is the transmission probability for an electron of energy $E$ to tunnel across the junction.  This transmission function may be expressed in terms of the junction's Green's functions as:\cite{Datta95}  
\begin{equation}
{\myT}(E)={\rm Tr}\left\{ \Gamma^1(E) G(E) \Gamma^2(E) G^\dagger(E)\right\},
\label{eq:transmission_prob}
\end{equation}
where $\Gamma^\alpha(E)$ is the tunneling-width matrix for lead $\alpha$
and $G(E)$ is the retarded Green's function of the SMJ.  

In organic molecules, such as those considered here, electron-phonon coupling is weak, allowing $ZT$ to be expressed as follows: 
\begin{equation}
	ZT = \left. ZT \right|_{el}\left(\frac{1}{1+\kappa^{ph}/\kappa^{el}}\right),
	\label{eq:ZT_full}
\end{equation}
where\cite{Finch09}
\begin{equation}
\left. ZT \right|_{el} = \left(\frac{\myL{0}\myL{2}}{\left[\myL{1}\right]^2}-1\right)^{-1}
\label{eq:ZT_in_L}
\end{equation}
and
\begin{equation}
\myL{\nu}\left(\mu,T\right) = \int dE  (E-\mu)^{\nu}\,\myT(E) \left(-\frac{\partial f_0}{\partial E}\right).
\label{eq:Lnu}	
\end{equation}
Here $f_0$ is the equilibrium Fermi function and $\kappa^{ph}$=$\kappa_0 \myT^{ph}$ is the phonon's thermal conductance, where $\kappa_0$=$(\pi^2/3)(k_{\rm B}^2 T/h)$ is the thermal conductance quantum\cite{Rego99} and $\myT^{ph}$ is the phonon transmission probability.  Since the Debye frequency in the metal lead is typically smaller than the lowest vibrational mode of a small organic molecule, the spectral overlap of phonon modes between the two is small, implying $\myT^{ph}$$\ll$1 and consequently that $ZT$$\approx$$\left.ZT\right|_{el}$.

\begin{figure}[htb]
	\centering
	\includegraphics[width=2.5in]{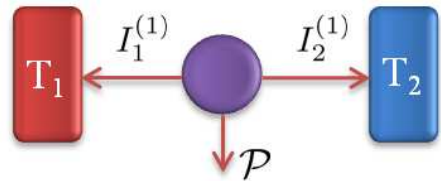}
	\captionsetup{singlelinecheck=off,justification=RaggedRight}
	\caption[Heat pump]{Schematic diagram of a thermoelectric device, where $I^{(1)}_\alpha$ is the heat current flowing into lead $\alpha$, $T_\alpha$ is the temperature and ${\gcal P}$ is the power output.  
	}
	\label{fig:thermo_diagram}
	\vspace{-.5cm}
\end{figure}

Thermodynamically, a system's response is characterized by the efficiency $\eta$ with which heat can be converted into usable power ${\gcal P}$ and the amount of power that can be generated.  Applying the first law of thermodynamics to the device shown in Fig.~(\ref{fig:thermo_diagram}) gives:
\begin{equation}
	{\gcal P}= - {I}^{(1)}_1-{I}^{(1)}_2  = I_1^{(0)} (\mu_1 - \mu_2),
	\label{eq:thermo_power}
\end{equation}
where 
we mention that the power is equivalently phrased in terms of heat or electrical currents.  The efficiency $\eta$ is defined as the ratio of power output to input heat current:
\begin{equation}
\eta=\frac{{\gcal P}}{\left|I^{(1)}_1\right|}=
	- \frac{I^{(1)}_1 + I^{(1)}_2}{\left|I^{(1)}_1\right|},
	\label{eq:thermo_eff}
\end{equation}
where we have assumed that $T_1>T_2$.  
With these expressions for the power and efficiency, we can completely quantify the performance of a quantum device, both near and far from equilibrium.

\begin{figure*}[tb]
	\centering
	\subfloat[many-body theory]{\label{fig:many-body_benzene}
	\put(135,162){\includegraphics[width=.65in]{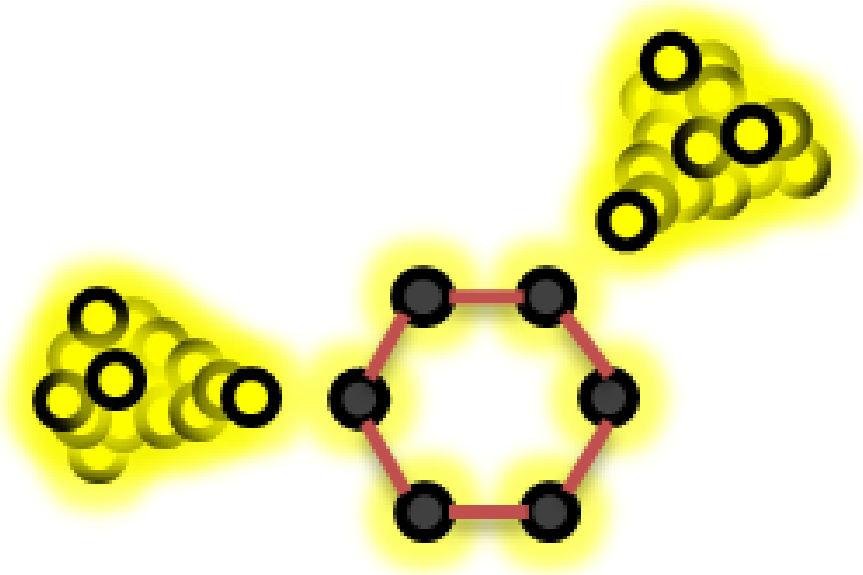}} 
	\includegraphics[width=.51\linewidth]{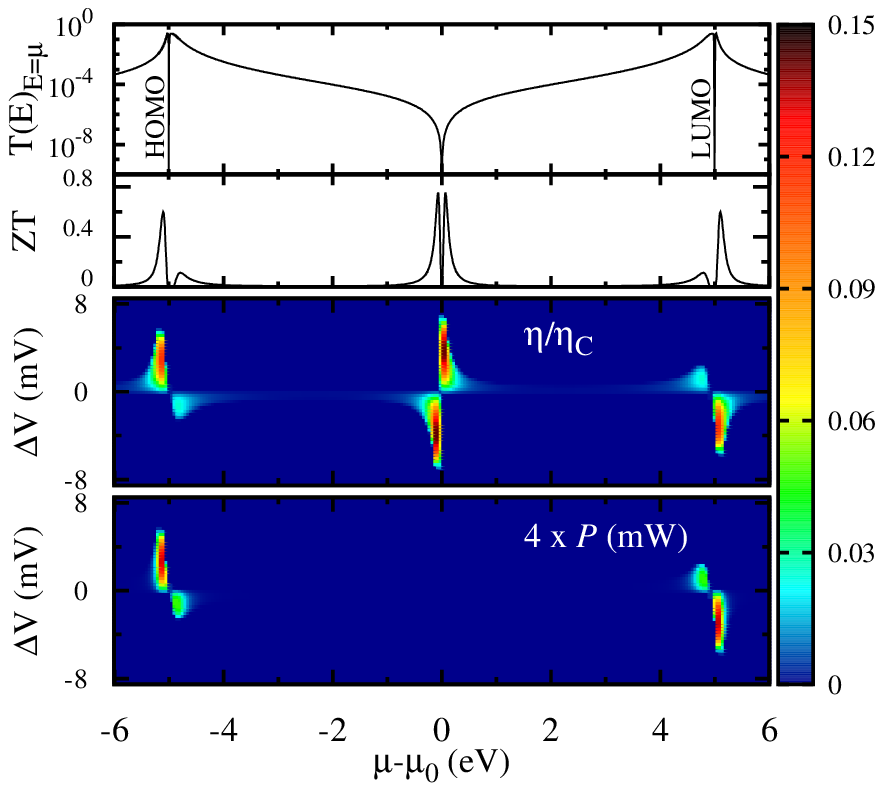}}
	\subfloat[H\"uckel theory]{\label{fig:Huckel_benzene}
	\includegraphics[width=.51\linewidth]{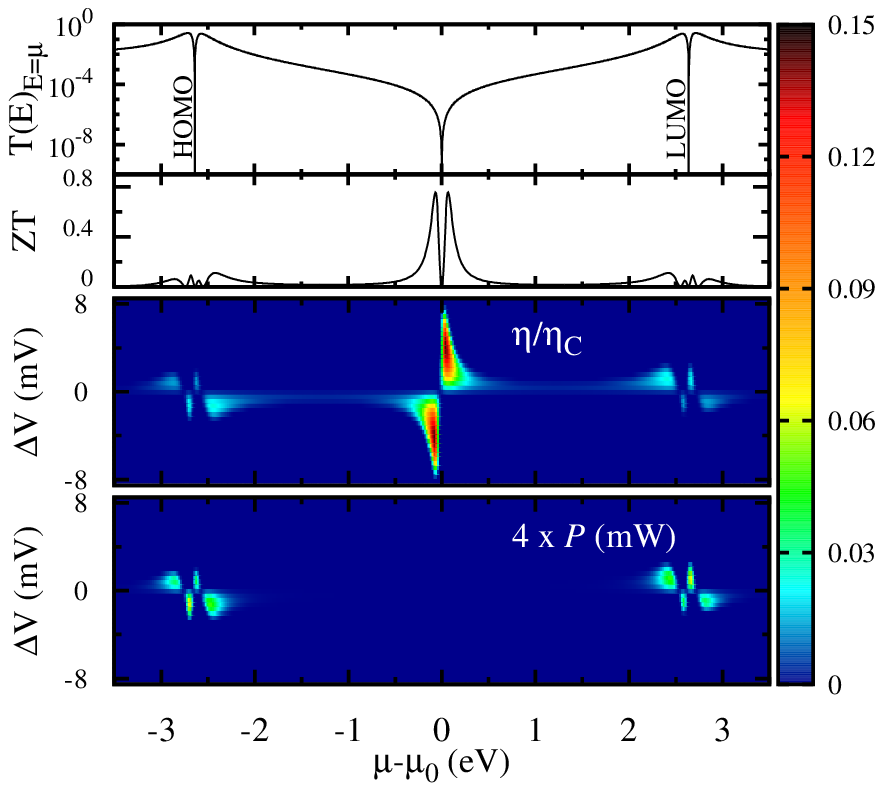}}
		\caption[Full-width many-body vs. H\"uckel]{The transmission probability $\myT(E)$,  figure-of-merit $ZT$, Carnot-normalized efficiency $\eta/\eta_{\rm C}$, and electrical power output ${\gcal P}$ of a two terminal 1,3-benzene SMJ, with lead temperatures $T_1$=300K and $T_2$=250K, calculated using (a) many-body and (b) H\"uckel theory, highlighting the discrepancies near resonances and the similarities near the node in the two theories.  As a function of $\mu$, $\eta$ and $ZT$ are in excellent qualitative agreement while ${\gcal P}$ is only peaked near resonance, suggesting that $ZT$ is incomplete as a device performance metric. (a) Many-body calculations give ${\gcal P}_{\rm peak}$=33$\mu$W and $\eta_{\rm peak}/\eta_{\rm C}$=11.5\% near resonance. (b)  H\"uckel calculations give ${\gcal P}_{\rm peak}$=21$\mu$W and $\eta_{\rm peak}/\eta_{\rm C}$=2.7\% near resonance.  The mid-gap region is discussed in Fig.~(\ref{fig:many-body_huckel_closeup}).  Note that the peak $ZT$=0.75 is on par with currently available commercial thermoelectrics.\cite{Snyder08,Bell08}  Calculations were performed using the model and parameterization of benzene discussed in detail in Ref.7 with $\Gamma$=0.63eV. 
\label{fig:benzene_figure3}
\vspace{-.5cm}
		}
\end{figure*}


As a first example, we calculate the non-linear thermodynamic response of a meta-connected Au-benzene-Au SMJ using many-body\cite{Bergfield09} and H\"uckel theory, shown in Fig.~(\ref{fig:many-body_benzene}) and Fig.~(\ref{fig:Huckel_benzene}), respectively.  Although the transmission spectrum of this junction doesn't possess a supernode, it does possess a quadratic node within $\pi$-electron theory,\cite{Bergfield09b, Cardamone06} and will allow us to ascertain the importance of interactions on the thermoelectric response of a SMJ.
%
%

In the top panel of each figure is a section of the transmission spectrum, showing the {\sc homo} and {\sc lumo} resonances and the quadratic node directly in between at $\mu$=$\mu_0$. Associated with this node is an enhancement in many linear-response metrics\cite{Bergfield09b} including $ZT$, which is shown in the second panel from the top.  The bottom two portions of each figure show the calculated efficiency $\eta$ and power ${\gcal P}$ when a junction with $T_1$=300K and $T_2$=250K is further pushed out of equilibrium via the application of a bias voltage $\Delta V$.  In all simulations presented here, the lead-molecule coupling is taken to be symmetric such that $\Gamma^\alpha_{nm}$=$\Gamma \delta_{na}\delta_{ma}$, where $n$, $m$, and $a$ are $\pi$-orbital labels and $a$ is coupled to lead $\alpha$.  The efficiency is normalized with respect to the maximum allowed by the second law of thermodynamics, the Carnot efficiency $\eta_{\rm C}=\Delta T /T_1$, where $\Delta T$=$T_1$-$T_2$. 

The nonequilibrium thermodynamic response of a 1,3-benzene SMJ calculated using many-body theory is shown in Fig.~(\ref{fig:many-body_benzene}).  The $ZT$ and $\eta$ spectra, shown in two middle panels of the same figure, exhibit peaks in the vicinity of both transmission nodes and resonances whereas the power ${\gcal P}$, shown in the bottom panel, is only peaked near transmission resonances.  Around either the {\sc homo} or {\sc lumo} resonance, the peak power ${\gcal P}_{\rm peak}$=33$\mu$W and peak efficiency $\eta_{\rm peak}/\eta_{\rm C}$=11.5\% are only realized when the junction operates out of equilibrium at a bias voltage $\Delta V$=3mV.  With a chemical potential near the mid-gap node and $\Delta V$=3.6mV $\eta_{\rm peak}/\eta_{\rm C}$=14.9\%, larger than near resonance but with a much lower peak power ${\gcal P}_{\rm peak}$=.088nW.

In the vicinity of a resonance, there are both quantitative and qualitative differences in the linear and non-linear thermodynamic response predicted by the two theories.
By neglecting interactions, the H\"uckel theory fails to accurately predict both the degeneracy and position of electronic resonances.  It also incorrectly determines the peak values of $ZT$, $\eta$ and ${\gcal P}$ in the vicinity of a resonance. As can be seen near either ({\sc homo} or {\sc lumo}) resonance in Fig.~(\ref{fig:benzene_figure3}), the H\"uckel theory predicts a Carnot-normalized peak efficiency of 2.7\% which is nearly five times less than the 11.5\% predicted by the many-body theory.  The peak power near a resonance also varies considerably between the two theories, where the H\"uckel calculations give ${\gcal P}_{\rm peak}$=21$\mu$W while many-body theory predicts ${\gcal P}_{\rm peak}$=33$\mu$W.  These results indicate that interactions are required to accurately predict the thermoelectric response of devices operating in the resonant-tunneling regime.  It is interesting to note, however, that in both models the linear-response metric $ZT$ qualitatively captures the features of the non-linear metric $\eta$.

\begin{figure}[b]
	\centering
	\subfloat[$ZT$]{\label{fig:closeup_a} \includegraphics[width=.333\mycolumnwidth]{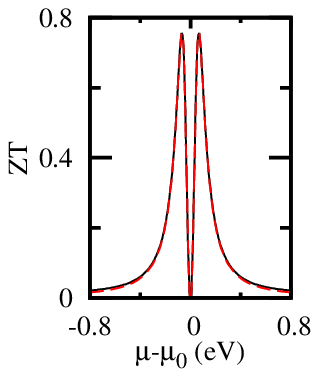}}
	\subfloat[$\eta/\eta_{\rm C}$]{\label{fig:closeup_b}\includegraphics[width=.333\mycolumnwidth]{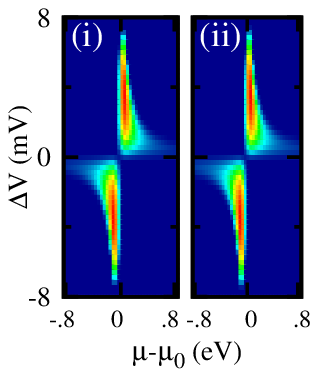}}
	\subfloat[50$\times {\gcal P}$ ($\mu$W)]{\label{fig:closeup_c}\includegraphics[width=.333\mycolumnwidth]{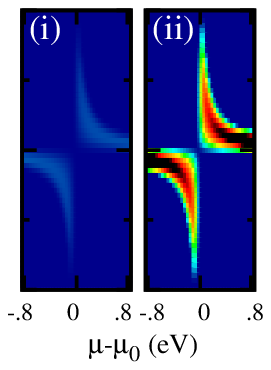}}
	\caption[Figure 4, Hueckel vs. many-body]{Calculations of $ZT$, $\eta$ and ${\gcal P}$ in the vicinity of the transmission node at $\mu$=$\mu_0$ of a meta-benzene SMJ using many-body (red line and panel i) and H\"uckel (black line and panel ii) theories.  (a) and (b): $ZT$ and $\eta$ are found to be identical and independent of theory. (c) ${\gcal P}$ is strongly affected by interactions where, at peak efficiency ($\eta_{\rm peak}/\eta_{\rm C}$=14.91\%), many-body and H\"uckel calculations give ${\gcal P}_{\rm max}$=.088nW and ${\gcal P}_{\rm max}$=1.87nW, respectively.  The simulation parameters and colorscale are the same as in Fig.~(\ref{fig:benzene_figure3}).
	}
	\label{fig:many-body_huckel_closeup}
\end{figure}

In this article, we are interested in thermoelectric enhancement near nodes 
far away from any resonances.  Although interactions are {\em required} in order to ensure the invariance of transport quantities under a global voltage shift (i.e.~gauge-invariance), near the particle-hole symmetric point the effect of interactions on the thermoelectric response should be small.  In panels a-b of Fig.~(\ref{fig:many-body_huckel_closeup}), a comparison of $ZT$ and $\eta$ using both many-body and H\"uckel theories is shown near $\mu_0$ for a 1,3-benzene SMJ.  Near this point, $ZT$ and $\eta$ are independent of theory employed.
In contrast, the power, shown in panel c of the same figure, exhibits an order of magnitude difference between the two theories.  This observation can be understood by noticing that the calculated {\sc homo-lumo} gap is $\approx$10eV using many-body theory (panel c-i) whereas it is only $\approx$5.5eV when interactions are neglected in the H\"uckel theory (panel c-ii).  Since the power is peaked near transmission resonances, whose widths are fixed by the lead-molecule coupling $\Gamma$, the larger gap found using many-body theory gives a correspondingly lower predicted power.  


While the H\"uckel theory is not able to accurately characterize the thermoelectric response of a junction in the resonant-tunneling regime, it is sufficient for predicting $\eta$ and $ZT$ in the vicinity of the transmission node.  Since we are interested in these quantities for mid-gap supernodes, we shall use H\"uckel theory to simulate the larger molecules presented below.

The transmission node in a meta-benzene junction can be understood in terms of destructive interference of electron waves 
traversing the ring at the Fermi energy.\cite{Cardamone06}
According to Luttinger's theorem,\cite{Langreth66, Luttinger60} the Fermi volume is unaffected by the inclusion of electron-electron interactions.  
Consequently, in an aromatic ring such as benzene 
the Fermi wavevector $k_{\rm F}$=$\pi/2d$ is conserved and is therefore sufficient to characterize quantum interference both with and without interactions near $\mu_0$, since $\Delta\phi$=$ k_{\rm F}\Delta l$, where $\Delta\phi$ is the relative phase between transport paths with length difference $\Delta l$, and $d$ is the inter-site distance.

\begin{figure}[b]
	\centering
		\begin{picture}(0,0)
	 \put(137,99){\includegraphics[width=.95in]{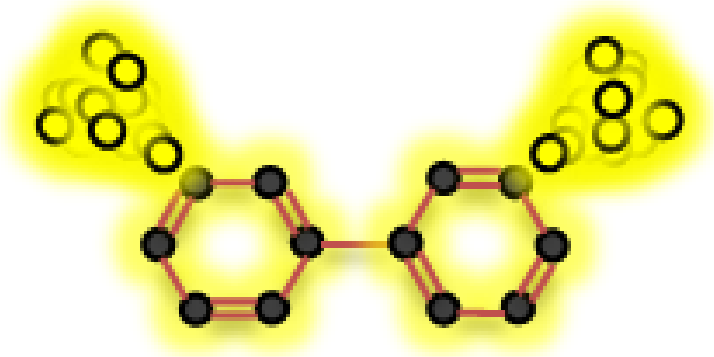}}
	 \end{picture} 
	\includegraphics[width=\mycolumnwidth]{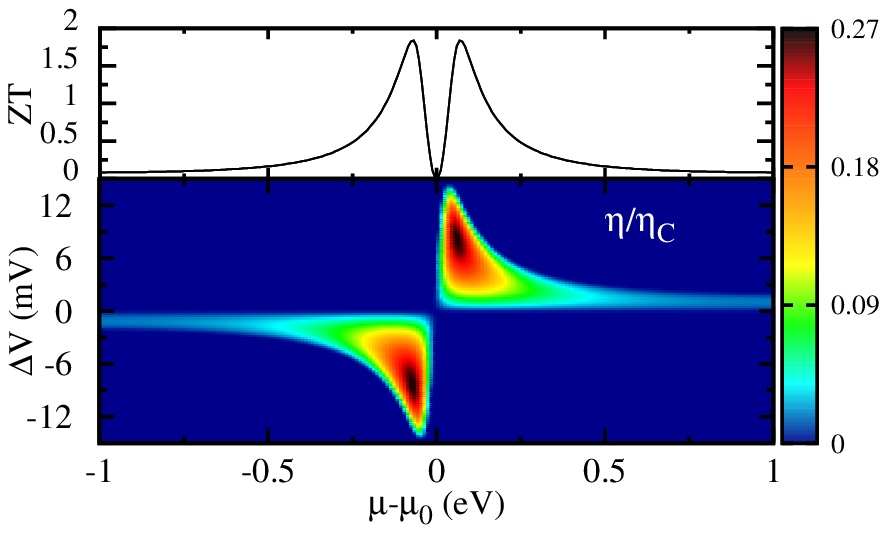} 
	\caption[Biphenyl $ZT$ and $\eta$]{A closeup of $ZT$ and $\eta$ near the quartic supernode of a 3,3'-biphenyl SMJ showing $ZT_{\rm peak}$=1.84 and $\eta_{\rm peak}/\eta_{\rm C}$=26.86\% at a predicted power of 0.75pW.  The junction geometry is shown schematically in the inset of the upper panel.  Simulations were performed using H\"uckel theory with $T_1$=300K, $T_2$=250K and $\Gamma$=0.5eV.}
\label{fig:Biphenyl}
\end{figure}

This is an important result, since the energy of resonant levels will generally depend strongly on whether or not interactions are included.  
Since $k_{\rm F}$ is protected, however, the transmission node across a single phenyl group 
is not so much a coincidence of energy levels as a {\em wave phenomenon}, meaning that interference in molecules composed of 
multiple aromatic rings in series
can be understood in terms of the interference within each subunit 
rather than the energy spectrum of the entire molecule.  
We find that such polycyclic molecules 
can exhibit higher-order {\em supernodes}, 
and that associated with a supernode is an order-dependent quantum enhancement of the junction's thermoelectric response.  Additional transport channels (e.g. $\sigma$-orbitals) or incoherent processes may lift the supernode.  The effect on the thermoelectric response is small provided the processes are weak, as discussed in Ref.~(\onlinecite{Bergfield09b}).

\begin{figure}[tb]
	\centering
\begin{center}
	 \includegraphics[width=3.1in]{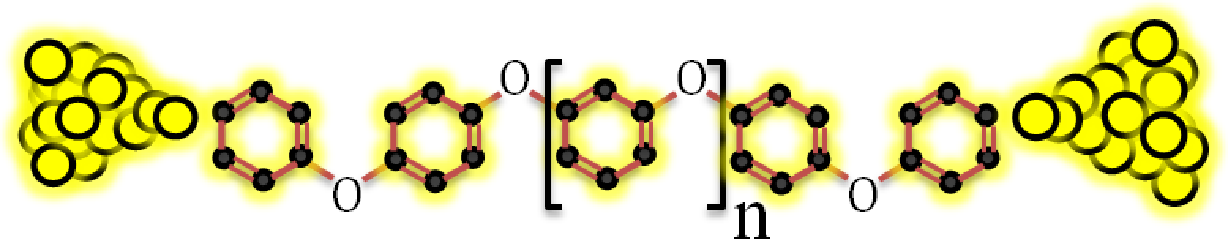}
	 \hspace{-1cm}	
	 \vspace{-.4cm}
\end{center}
	\includegraphics[width=\mycolumnwidth]{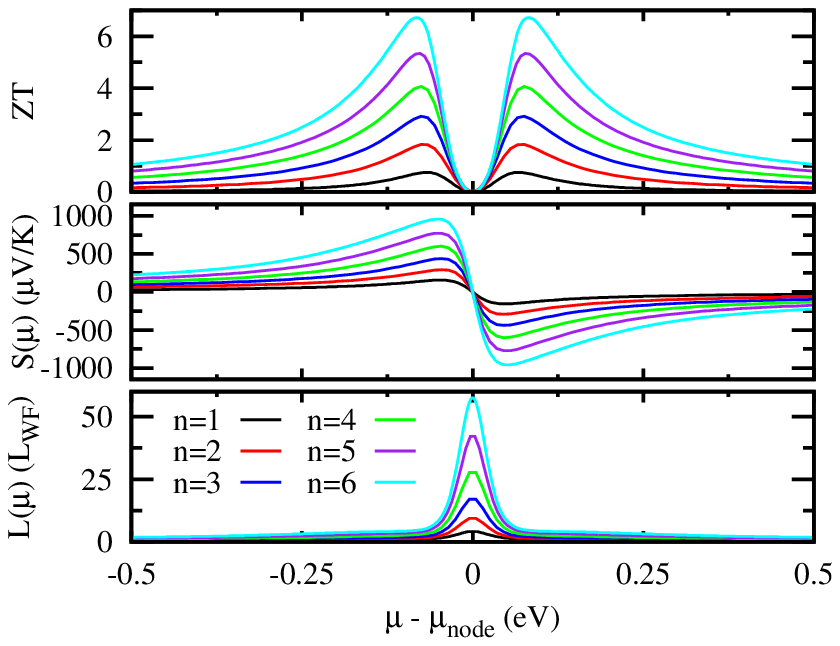} 
	\caption[Supernode triplot]{Supernode enhancement of $ZT$, thermopower $S$ and Lorenz number $L$ for polyphenyl ether (PPE) SMJs with $n$ repeated phenyl groups, shown schematically above the top panel.  As a function of $n$, $ZT_{\rm peak}$ scales super-linearly exhibiting a peak value of 6.86 for $n$=6.  The thermopower and Lorenz number are also enhanced with $S_{\rm peak}$=957$\mu$V/K and $L_{\rm peak}$=55.33$L_{\rm WF}$ at the same value of $n$.  Simulations were performed using H\"uckel theory at room temperature ($T$=300K) with $\Gamma$=0.5eV.  Inter-phenyl electronic hopping was set an order of magnitude below the intra-phenyl value of 2.64eV.}
\label{fig:supernode_triplot}
\end{figure}

The 3,3'-biphenyl junction, drawn schematically in the top panel of Fig.~(\ref{fig:Biphenyl}), can viewed as two meta-connected benzene rings in series.  This junction
geometry is similar to that studied by Mayor et al.\cite{Mayor03}  In agreement with the prediction that a biphenyl junction should possess a quartic supernode, the linear and non-linear response shown in Fig.~(\ref{fig:Biphenyl}) exhibits peak values of efficiency ($\eta/\eta_{\rm C}$=26.86\%) and $ZT$ (1.84) that are over twice those of benzene.  With $ZT$$\approx$2, the biphenyl junction exhibits sufficient thermoelectric performance to be attractive for many commerical solid-state heating and cooling applications.\cite{Bell08, DiSalvo99, Snyder08}  As we shall see, this is only the first in an entire class of supernode-possesing molecules which exhibit even larger values of $\eta$ and $ZT$.

In larger molecules composed of $n$ meta-connected phenyl group in series, we expect that the transmission nodes should combine and give rise to a 2$n$$^{\rm th}$ order supernode.  Polyphenyl ether (PPE), shown schematically at the top of Fig.~(\ref{fig:supernode_triplot}), consists of $n$ phenyl rings connected in series with ether linkages.  Based on our previous discussion, we predict that a PPE-based junction should exhibit a 2$n$$^{th}$ order supernode.  The figure-of-merit $ZT$, thermopower $S$ and Lorenz number $L$=$\kappa/GT$ for PPE junctions are shown in the top, middle and bottom panels of Fig.~(\ref{fig:supernode_triplot}), respectively, where the Lorenz number is normalized with respect to the Wiedemann--Franz (WF) value L$_{\rm WF}$=$\left(\pi^2/3\right)(k_{\rm B}/e)^2$.

The bottom panel of Fig.~(\ref{fig:supernode_triplot}) shows an increasing peak Lorenz number $L_{\rm peak}$ with increasing $n$.  In linear-response,  $L$ and $S$ can be expressed in terms of Eq.~(\ref{eq:Lnu}) as:
\begin{equation}
\left. L\right|_{el}  = \frac{1}{(eT)^2} \left(\frac{\myL{2}}{\myL{0}} - \left[\frac{\myL{1}}{\myL{0}} \right]^2 \right),
\label{eq:Lorenz}
\end{equation}
and $S$=$-\frac{1}{eT} \frac{\myL{1}}{\myL{0}}$, where $e$ is the magnitude of the electron's charge and $T$ is the temperature.  Using Eq.~(\ref{eq:Lorenz}) and Eq.~(\ref{eq:Lnu}) with the transmission function of Eq.~(\ref{eq:T_supernode}) we find that:
\begin{equation}
	\left. \frac{L_{\rm max}}{L_{\rm WF}} \right|_{el} = \left( \frac{3}{\pi^2} \right)\frac{\left. \left[ \partial_b^{2n+2} b\pi \csc(b\pi)\right] \right|_{b=0}}{\left. \left[ \partial_b^{2n}  b\pi \csc(b\pi)  \right] \right|_{b=0}}.
	\label{eq:Lmax_analytic}
\end{equation}
Setting $n$=6 in Eq.~(\ref{eq:Lmax_analytic}) gives $L_{\rm max}$=55.33$L_{\rm WF}$, corresponding exactly to the result of the full simulation shown in the bottom panel of Fig.~(\ref{fig:supernode_triplot}).  Similar agreement is found for the other values of $n$, confirming the presence of 2n$^{th}$ order supernodes in these junctions.

We find that higher-order supernodes in the transmission spectrum of a nanoscale junction give rise to an order-dependent quantum-enhancement of the linear and non-linear thermoelectric response.  The full nonequilibrium spectrum of thermodynamic efficiency qualitatively resembles the figure-of-merit $ZT$ spectrum, suggesting that $ZT$ encapsulates the salient physics related to efficiency even at the nanoscale.  Efficiency, however, is only part of a device's performance.  Another important quantity is the usable power produced by a device, whose variations are poorly characterized by $ZT$ at the nanoscale.  


Thermoelectric devices based on individual SMJs are ideally suited for local cooling in integrated nanoscale circuit architectures.  Supernode-based 
devices have a low transmission probability and thus a large electrical impedance capable of withstanding voltage surges.
Moreover, high-power macroscopic devices could be constructed by growing layers of densely packed 
molecules.  For example, a self-assembled
monolayer with a surface density\cite{Zangmeister04} 
of 4$\times$10$^{15}$molecules/cm$^2$ would give 352kW/cm$^2$ at peak efficiency for a meta-benzene film.  
The efficiency of PPE-based devices increases with ring number and is only limited by the electronic coherence length, suggesting that highly efficient molecular-based thermoelectric devices may soon be realized.


 
\bibliography{refs}

\end{document}